\documentclass[twocolumn,showpacs,preprintnumbers,amsmath,amssymb,pre]{revtex4}

\usepackage{epsfig}
\usepackage{bm}

\begin{document}


\newcommand{\dd}{\mathrm{d}}
\newcommand{\gcc}{\mbox{g~cm$^{-3}$}}
\newcommand{\kB}{k_\mathrm{B}}
\newcommand{\etal}{{et al.}}
\newcommand{\beq}{\begin{equation}}
\newcommand{\eeq}{\end{equation}}
\newcommand{\bea}{\begin{eqnarray}}
\newcommand{\eea}{\end{eqnarray}}
\newcommand{\req}[1]{Eq.\ (\ref{#1})}
\newcommand{\Znuc}{Z_\mathrm{nuc}}
\newcommand{\Ftot}{F_\mathrm{tot}}


\newcommand{\ApJ}[1]{{Astrophys.\ J.} \textbf{#1}}
\newcommand{\ApJS}[1]{{Astrophys.\ J. Suppl.\ Ser.} \textbf{#1}}
\newcommand{\AandA}[1]{{Astron.\ Astrophys.} \textbf{#1}}
\newcommand{\ARAA}[1]{{Annu.\ Rev.\ Astron.\ Astrophys.} \textbf{#1}}
\newcommand{\PR}[1]{{Phys.\ Rev.} \textbf{#1}}
\newcommand{\PRA}[1]{{Phys.\ Rev. A} \textbf{#1}}
\newcommand{\PRE}[1]{{Phys.\ Rev. E} \textbf{#1}}
\newcommand{\PRL}[1]{{Phys.\ Rev.\ Lett.} \textbf{#1}}


\title{Comments on ``Plasma oscillations and nonextensive statistics''}

\author{Xiao-Chang Chen}\email{xcchen1985@sina.com}%
\affiliation{School of Materials Science and Engineering, Nanchang
University, Nanchang 330031, China}
     \author{Xiao-Qing Li}\email{njxqli2010@163.com}
\affiliation{Department of Physics, Nanjing Normal University ,
Nanjing 210097, China}

\received{}

\begin{abstract}
The paper, authored by J. A. S. Lima et al, was published in
\textit{Phys. Rev. E} in 2000 has discussed the dispersion relation
and Landau damping of Langmuir wave in the context of the
nonextensive statistics proposed by Tsallis. It has been cited by
many authors because the dispersion relation in Tsallis formalism
present a good fit to the experimental data when $q < 1$, while the
classical result based on Maxwellian distribution only provides a
crude description. However, the results obtained in this paper are
problematic. In this comments on the paper we shall derive the
correct analytic formulas both for the dispersion relation and
Landau damping in Tsallis formalism. We hope that this comments will
be useful in providing the correct results.

\end{abstract}



\maketitle

\section{Introduction}

Over the last few years, it has been proven that systems which
present long-range interactions, long-time memory, fractality of the
corresponding space-time, or intrinsic inhomogeneity are intractable
within the conventional Boltzmann-Gibbs statistics [1, 2]. So there
has been an increasing focus on the new statistical approach, i.e.
nonextensive statistical mechanics (NSM), in recent years. For $q
\ne 1$, it gives power-law distribution and only when the parameter
$q \to 1$ Maxwellian distribution is recovered[3]. NSM has been
successfully applied to stellar polytropes [4], two dimensional
Euler and drift turbulence in a pure electron plasma column [5], as
well as to the peculiar velocity function of galaxies clusters [6].
In particular, Liu et al. [7] showed a reasonable indication for the
non-Maxwellian velocity distribution from plasma experiments.

Dispersion relations are fundamental and important for studying the
wave in the plasma. According to the dispersion relations, we can
study the problem of instability, propagation, refraction and
absorption of the plasma wave. Recently, there has been a great deal
of interest in studying the dispersion property in plasmas in the
context of the nonextensive statistics. The paper authored by J. A.
S. Lima et al [8], has a considerable hold over the field of plasma
physics, have studied the dispersion relation of Langmuir wave based
on q-distribution, the results show that nonextensive formalism
presents a good fit to the experimental data, while the standard
Maxwellian distribution only provides a crude description. However,
the results obtained in this paper are problematic. The reason is
that the author investigates the propagation of electrostatic waves
by using the one-dimensional equilibrium distribution. But the
equilibrium distribution in the dielectric function should be
marginal distribution (quasi one-dimensional distribution), which is
different with one-dimensional distribution in the context of
nonextensive statistics unlike the classical Boltzmann-Gibbs
statistics. In this comments on the paper we will derive the correct
analytic formulas both for the dispersion relation and Landau
damping in detail with Tsallis formalism. It is our hope that the
discussion here will be useful in the field of plasma physics.

The paper is organized as follows. In Section II we briefly
introduce the nonextensive distribution function. The generalized
dispersion relation and Landau damping for Langmuir wave are
obtained in Section III. Finally, the summary is given in Section
IV.

\section{Nonextensive distribution function}
\label{sect-fren} First let us recall some basic facts about Tsallis
statistics. In Tsallis statistics, the entropy has the form [3] of

\begin{equation}
\label{eq1} S_q = k_B \frac{1 - \sum\nolimits_i {p_i^q } }{q - 1},
\end{equation}

\noindent where $k_B $ is the Boltzmann constant, $q$ is a parameter
quantifying the degree of nonextensivity, $p_i $ is the probability
of the $i$th microstate. The B-G entropy is recovered in the limit
$q \to 1$. The basic property of Tsallis entropy is the
nonadditivity or nonextensivity for $q \ne 1$. For example, for two
systems A and B, the rule of composition [3] reads

\begin{equation}
\label{eq2} S_q (A + B) = S_q (A) + S_q (B) + (1 - q)S_q (A)S_q (B).
\end{equation}

In the nonextensive description, the three-dimensional equilibrium
distribution function can be written as [9]

\begin{equation}
\label{eq3} f_q \left( {\rm {\bf p}} \right) = A_q [1 - \left( {q -
1} \right)\frac{{\rm {\bf p}}^2}{2m^2v_T^2 }]^{\frac{1}{q - 1}},
\end{equation}

\noindent according to the normalizing condition

\begin{equation}
\label{eq4} \int {f_q \left( {\rm {\bf p}} \right)d{\rm {\bf
p}}\frac{1}{\left( {2\pi } \right)^3}} = n_0 ,
\end{equation}

\noindent the normalization constant reads

\begin{equation}
\label{eq5} A_q = L_q \frac{\left( {\sqrt {2\pi } }
\right)^3}{\left( {mv_T } \right)^3}n_0 ,
\end{equation}

\noindent in which

\begin{equation}
\label{eq6} L_q = \frac{\Gamma \left( {\frac{1}{1 - q}}
\right)}{\left( {\frac{1}{1 - q}} \right)^{3 \mathord{\left/
{\vphantom {3 2}} \right. \kern-\nulldelimiterspace} 2}\Gamma \left(
{\frac{1}{1 - q} - \frac{3}{2}} \right)}, \quad \frac{1}{3} < q \le
1
\end{equation}

\noindent and

\begin{equation}
\label{eq7} L_q = \frac{3q - 1}{2}\frac{\left( {\frac{1}{q - 1}}
\right)^{{ - 3} \mathord{\left/ {\vphantom {{ - 3} 2}} \right.
\kern-\nulldelimiterspace} 2}\Gamma \left( {\frac{1}{q - 1} +
\frac{3}{2}} \right)}{\Gamma \left( {\frac{1}{q - 1}} \right)},
\quad q \ge 1
\end{equation}

\noindent where ${\rm {\bf p}}$, $v_T = \sqrt {{k_B T}
\mathord{\left/ {\vphantom {{k_B T} m}} \right.
\kern-\nulldelimiterspace} m} $, $k_B $, $T$, $m$, and $n_0 $
denotes, respectively, the momentum of particles, thermal speed,
Boltzmann constant, temperature of particles, mass of particles and
particle number density. As one may check, for $q < 1
\mathord{\left/ {\vphantom {1 3}} \right. \kern-\nulldelimiterspace}
3$, the q-distribution is unnormalizable. For $1 \mathord{\left/
{\vphantom {1 3}} \right. \kern-\nulldelimiterspace} 3 < q \le 1$,
the momentum of the particles can take any value. For $q \ge 1$, the
distribution function Eq.(\ref{eq3}) exhibits a cutoff on the
maximum value allowed for the momentum of the particles, which is
given by

\begin{equation}
\label{eq8} p_{\max } = \sqrt {2 \mathord{\left/ {\vphantom {2
{\left( {q - 1} \right)}}} \right. \kern-\nulldelimiterspace}
{\left( {q - 1} \right)}} mv_T ,
\end{equation}

We see that in the limit $q \to 1$, $p_{\max } $ goes to infinity
and Eq.(\ref{eq3}) reduces to the Boltzmann distribution function

\begin{equation}
\label{eq9} f_{q = 1} \left( {\rm {\bf p}} \right) = \frac{\left(
{\sqrt {2\pi } } \right)^3}{\left( {mv_T } \right)^3}n_0 \exp ( -
\frac{{\rm {\bf p}}^2}{2m^2v_T^2 }).
\end{equation}

In order to define the temperature of the system which is described
by the nonextensive distribution, we will calculate the average
kinetic energy below. For $1 \mathord{\left/ {\vphantom {1 3}}
\right. \kern-\nulldelimiterspace} 3 < q \le 1$,

\[
\left\langle {E_q } \right\rangle = \left\langle {\frac{{\rm {\bf
p}}^2}{2m}} \right\rangle = \int {\frac{{\rm {\bf p}}^2}{2m}f_q
\left( {\rm {\bf p}} \right)d{\rm {\bf p}}\frac{1}{\left( {2\pi }
\right)^3}}
\]

\[
 = \frac{L_q }{\left( {2\pi } \right)^3}\frac{\left( {\sqrt {2\pi } }
\right)^3}{\left( {mv_T } \right)^3}n_0 \cdot
\]

\[
\int_0^\infty {\frac{p^2}{2m}[1 - \left( {q - 1}
\right)\frac{p^2}{2m^2v_T^2 }]^{\frac{1}{q - 1}}4\pi p^2dp}
\]

\[
 = \frac{L_q }{\sqrt {2\pi } }\frac{1}{\left( {mv_T } \right)^3}\frac{n_0
}{m}\int_0^\infty {p^4[1 + \frac{1 - q}{2m^2v_T^2 }p^2]^{ -
\frac{1}{1 - q}}dp}
\]

\begin{equation}
\label{eq10}
 = \frac{2}{5q - 3}\frac{3}{2}n_0 mv_T^2 = \frac{2}{5q - 3}\frac{3}{2}n_0
k_B T.
\end{equation}

\noindent where Eq.(\ref{eq10}) has been calculated using the
integral formula [10, p325], that is,

\begin{equation}
\label{eq11} \int_0^\infty {x^{\mu - 1}\left( {1 + \beta x^p}
\right)^{ - \nu }dx = \frac{1}{p}\beta ^{ - \frac{\mu }{p}}B\left(
{\frac{\mu }{p},\nu - \frac{\mu }{p}} \right)}
\end{equation}

\noindent with $\left| {\arg \beta } \right| < \pi ,p > 0,0 < Re\mu
< pRe\nu $, here require $q > 3 \mathord{\left/ {\vphantom {3 5}}
\right. \kern-\nulldelimiterspace} 5$. $B$ is the Beta function, the
relation of Beta function and Gamma function is [10, p909]

\begin{equation}
\label{eq12} B\left( {x,y} \right) = \frac{\Gamma \left( x
\right)\Gamma \left( y \right)}{\Gamma \left( {x + y} \right)}.
\end{equation}

For $q \ge 1$,

\[
\left\langle {E_q } \right\rangle = \left\langle {\frac{{\rm {\bf
p}}^2}{2m}} \right\rangle = \int {\frac{{\rm {\bf p}}^2}{2m}f_q
\left( {\rm {\bf p}} \right)d{\rm {\bf p}}\frac{1}{\left( {2\pi }
\right)^3}}
\]

\[
 = \frac{L_q }{\left( {2\pi } \right)^3}\frac{\left( {\sqrt {2\pi } }
\right)^3}{\left( {mv_T } \right)^3}n_0 \cdot
\]

\[
\int_0^{p_{\max } } {\frac{p^2}{2m}[1 - \left( {q - 1}
\right)\frac{p^2}{2m^2v_T^2 }]^{\frac{1}{q - 1}}4\pi p^2dp}
\]

\[
 = \frac{L_q }{\sqrt {2\pi } }\frac{1}{\left( {mv_T } \right)^3}\frac{n_0
}{m}\int_0^{p_{\max } } {p^4[1 - \left( {q - 1}
\right)\frac{p^2}{2m^2v_T^2 }]^{\frac{1}{q - 1}}dp}
\]

\[
 = \frac{L_q n_0 }{\sqrt \pi }\frac{2mv_T^2 }{\left( {q - 1} \right)^{5
\mathord{\left/ {\vphantom {5 2}} \right. \kern-\nulldelimiterspace}
2}}\int_0^1 {t^{3 \mathord{\left/ {\vphantom {3 2}} \right.
\kern-\nulldelimiterspace} 2}\left( {1 - t} \right)^{\frac{1}{q -
1}}dt}
\]

\begin{equation}
\label{eq13}
 = \frac{2}{5q - 3}\frac{3}{2}n_0 mv_T^2 = \frac{2}{5q - 3}\frac{3}{2}n_0
k_B T.
\end{equation}

\noindent where Eq.(\ref{eq13}) has been calculated using the
transformation $t = {\left( {q - 1} \right)p^2} \mathord{\left/
{\vphantom {{\left( {q - 1} \right)p^2} {2m^2v_T^2 }}} \right.
\kern-\nulldelimiterspace} {2m^2v_T^2 }$ and integral formula [10,
p324], that is,

\begin{equation}
\label{eq14} \int_0^1 {x^{\mu - 1}\left( {1 - x^\lambda }
\right)^{\nu - 1}dx = \frac{1}{\lambda }B\left( {\frac{\mu }{\lambda
},\nu } \right)}
\end{equation}

\noindent with $Re\mu > 0,Re\nu > 0,\lambda > 0$. So the average
kinetic energy can be expressed as

\begin{equation}
\label{eq15} \left\langle {E_q } \right\rangle = \left\langle
{\frac{{\rm {\bf p}}^2}{2m}} \right\rangle = \frac{2}{5q -
3}\frac{3}{2}n_0 k_B T = \frac{3}{2}n_0 k_B T_q ,
\end{equation}

\noindent where $T_q = {2T} \mathord{\left/ {\vphantom {{2T} {\left(
{5q - 3} \right)}}} \right. \kern-\nulldelimiterspace} {\left( {5q -
3} \right)}$ is the physical temperature of the nonextensive system.
We see that in the limit $q \to 1$, $T_{q = 1} = T$ and the average
kinetic energy reduces to $\left\langle {E_{q = 1} } \right\rangle =
{3n_0 k_B T} \mathord{\left/ {\vphantom {{3n_0 k_B T} 2}} \right.
\kern-\nulldelimiterspace} 2$, which is the standard result in B-G
statistics.

\section{The generalized dispersion relation and Landau damping}
\label{sect-TDE}

For the longitudinal wave propagating in an unmagnetized,
collisionless, isotropic plasma, the longitudinal dielectric
function of electron can be written as [11]

\begin{equation}
\label{eq16} \varepsilon _k^l = 1 + \frac{4\pi e^2}{k^2}\int
{\frac{1}{\omega - {\rm {\bf k}} \cdot {\rm {\bf v}} + i\delta
}\left( {{\rm {\bf k}} \cdot \frac{\partial f_q \left( {\rm {\bf p}}
\right)}{\partial {\rm {\bf p}}}} \right)\frac{d{\rm {\bf
p}}}{\left( {2\pi } \right)^3}} ,
\end{equation}

\noindent we consider the direction of wave vector ${\rm {\bf k}}$
to be along x-axis, Eq.(\ref{eq16}) becomes

\[
\varepsilon _k^l = 1 + \frac{4\pi e^2}{k^2}\int {\frac{dp_x }{2\pi
}\frac{k\frac{\partial }{\partial p_x }}{\omega - kv_x + i\delta
}\int {f_q \left( {\rm {\bf p}} \right)\frac{dp_y dp_z }{\left(
{2\pi } \right)^2}} }
\]

\begin{equation}
\label{eq17}
 = 1 + \frac{4\pi e^2}{k^2}\int {\frac{k}{\omega - kv_x + i\delta
}\frac{\partial f_q \left( {p_x } \right)}{\partial p_x }\frac{dp_x
}{2\pi }} ,
\end{equation}

\noindent where $e$ is the electron charge, $i\delta $ comes from
Landau rules($\delta \to 0^ + )$ [12]. Note that $f_q \left( {p_x }
\right)$ is the marginal distribution in the nonextensive framework,
which is given by

\begin{equation}
\label{eq18} f_q \left( {p_x } \right) = \int {f_q \left( {\rm {\bf
p}} \right)\frac{dp_y dp_z }{\left( {2\pi } \right)^2}} .
\end{equation}

Next we will derive the expression of the marginal distribution.
Substituting Eq.(\ref{eq3}) into Eq.(\ref{eq18}), for $3
\mathord{\left/ {\vphantom {3 5}} \right. \kern-\nulldelimiterspace}
5 < q \le 1$, we obtain

\[
f_q \left( {p_x } \right) = \frac{4L_q }{\left( {mv_T }
\right)^3}\frac{n_0 }{\sqrt {2\pi } } \cdot
\]

\begin{equation}
\label{eq19} \int_0^\infty {dp_z \int_0^\infty {[1 - \left( {q - 1}
\right)\frac{p_x^2 + p_y^2 + p_z^2 }{2m^2v_T^2 }]^{\frac{1}{q -
1}}dp_y } } ;
\end{equation}

\noindent then the integral in Eq.(\ref{eq19}) over $p_y $ is

\[
\int_0^\infty {[1 - \left( {q - 1} \right)\frac{p_x^2 + p_y^2 +
p_z^2 }{2m^2v_T^2 }]^{\frac{1}{q - 1}}dp_y }
\]

\[
 = \int_0^\infty {[\frac{2m^2v_T^2 + \left( {1 - q} \right)\left( {p_x^2 +
p_z^2 } \right)}{2m^2v_T^2 } + \frac{\left( {1 - q}
\right)}{2m^2v_T^2 }p_y^2 ]^{ - \frac{1}{1 - q}}dp_y }
\]

\[
 = \left\{ {\frac{2m^2v_T^2 + \left( {1 - q} \right)\left( {p_x^2 + p_z^2 }
\right)}{2m^2v_T^2 }} \right\}^{ - \frac{1}{1 - q}} \cdot
\]

\[
\int_0^\infty {\left\{ {1 + \frac{\left( {1 - q} \right)}{2m^2v_T^2
+ \left( {1 - q} \right)\left( {p_x^2 + p_y^2 } \right)}p_y^2 }
\right\}^{ - \frac{1}{1 - q}}dp_y }
\]

\[
 = \frac{\sqrt \pi }{2}\frac{\Gamma \left( {\frac{1}{1 - q} - \frac{1}{2}}
\right)\left( {\frac{1}{1 - q}} \right)^{1 \mathord{\left/
{\vphantom {1 2}} \right. \kern-\nulldelimiterspace} 2}}{\Gamma
\left( {\frac{1}{1 - q}} \right)}\left( {2m^2v_T^2 }
\right)^{\frac{1}{1 - q}} \cdot
\]

\begin{equation}
\label{eq20} \left\{ {2m^2v_T^2 + \left( {1 - q} \right)\left(
{p_x^2 + p_z^2 } \right)} \right\}^{ - \frac{1}{1 - q} +
\frac{1}{2}},
\end{equation}

\noindent where Eq.(\ref{eq20}) has been calculated using the
integral formula (\ref{eq11}), and substituting Eq.(\ref{eq20}) into
Eq.(\ref{eq19}), according to the same method we can calculate the
integral over $p_z $. Finally Eq.(\ref{eq19}) becomes

\begin{equation}
\label{eq21} f_q \left( {p_x } \right) = \frac{L_q }{q}\frac{\sqrt
{2\pi } n_0 }{mv_T }[1 - \left( {q - 1} \right)\frac{p_x^2
}{2m^2v_T^2 }]^{\frac{1}{q - 1} + 1}.
\end{equation}

For $q \ge 1$, Substituting Eq.(\ref{eq3}) into Eq.(\ref{eq18}), we
obtain

\[
f_q \left( {p_x } \right) = \frac{4L_q }{\left( {mv_T }
\right)^3}\frac{n_0 }{\sqrt {2\pi } } \cdot
\]

\begin{equation}
\label{eq22} \int_0^{p_{z\max } } {dp_z \int_0^{p_{y\max } } {[1 -
\left( {q - 1} \right)\frac{p_x^2 + p_y^2 + p_z^2 }{2m^2v_T^2
}]^{\frac{1}{q - 1}}dp_y } } ,
\end{equation}

\noindent then the integral in Eq.(\ref{eq22}) over $p_y $ becomes

\[
\int_0^{p_{y\max } } {[1 - \left( {q - 1} \right)\frac{p_x^2 + p_y^2
+ p_z^2 }{2m^2v_T^2 }]^{\frac{1}{q - 1}}dp_y }
\]

\[
 = \int_0^{p_{y\max } } {[\frac{2m^2v_T^2 - \left( {q - 1} \right)\left(
{p_x^2 + p_z^2 } \right)}{2m^2v_T^2 } - }
\]

\[
\frac{\left( {q - 1} \right)}{2m^2v_T^2 }p_y^2 ]^{\frac{1}{q -
1}}dp_y
\]

\[
 = \left( {\frac{2m^2v_T^2 - \left( {q - 1} \right)\left( {p_x^2 + p_z^2 }
\right)}{2m^2v_T^2 }} \right)^{\frac{1}{q - 1}} \cdot
\]

\[
\int_0^{p_{y\max } } {\left( {1 - \frac{\left( {q - 1}
\right)}{2m^2v_T^2 - \left( {q - 1} \right)\left( {p_x^2 + p_y^2 }
\right)}p_y^2 } \right)^{\frac{1}{q - 1}}dp_y }
\]

\[
 = \frac{\left( {2m^2v_T^2 - \left( {q - 1} \right)\left( {p_x^2 + p_z^2 }
\right)} \right)^{\frac{1}{q - 1} + \frac{1}{2}}}{2\sqrt {q - 1}
\left( {2m^2v_T^2 } \right)^{\frac{1}{q - 1}}} \cdot
\]

\[
\int_0^1 {t^{ - \frac{1}{2}}\left( {1 - t} \right)^{\frac{1}{q -
1}}dt}
\]

\[
 = \frac{\sqrt \pi }{2}\frac{\Gamma \left( {\frac{1}{q - 1}} \right)}{\left(
{q - 1} \right)^{\frac{3}{2}}\Gamma \left( {\frac{1}{q - 1} +
\frac{3}{2}} \right)} \cdot
\]

\begin{equation}
\label{eq23} \frac{\left( {2m^2v_T^2 - \left( {q - 1} \right)\left(
{p_x^2 + p_z^2 } \right)} \right)^{\frac{1}{q - 1} +
\frac{1}{2}}}{\left( {2m^2v_T^2 } \right)^{\frac{1}{q - 1}}}
\end{equation}

\noindent where Eq.(\ref{eq23}) has been calculated using the
transformation $t = {\left( {q - 1} \right)p_y^2 } \mathord{\left/
{\vphantom {{\left( {q - 1} \right)p_y^2 } {\left( {2m^2v_T^2 -
\left( {q - 1} \right)\left( {p_x^2 + p_y^2 } \right)} \right)}}}
\right. \kern-\nulldelimiterspace} {\left( {2m^2v_T^2 - \left( {q -
1} \right)\left( {p_x^2 + p_y^2 } \right)} \right)}$ and integral
formula (\ref{eq14}), then substituting Eq.(\ref{eq23}) into
Eq.(\ref{eq22}), according to the same method we can calculate the
integral over $p_z $. Finally Eq.(\ref{eq22}) becomes

\begin{equation}
\label{eq24} f_q \left( {p_x } \right) = \frac{L_q }{q}\frac{\sqrt
{2\pi } n_0 }{mv_T }[1 - \left( {q - 1} \right)\frac{p_x^2
}{2m^2v_T^2 }]^{\frac{1}{q - 1} + 1}.
\end{equation}

Obviously, the marginal distribution Eqs.(\ref{eq21}) and
(\ref{eq24}) are different with one-dimensional distribution in the
context of nonextensive statistics [9]

\begin{equation}
\label{eq25} f_q \left( {p_x } \right) = B_q \frac{\sqrt {2\pi } n_0
}{mv_T }[1 - \left( {q - 1} \right)\frac{p_x^2 }{2m^2v_T^2
}]^{\frac{1}{q - 1}}
\end{equation}

\noindent in which

\begin{equation}
\label{eq26} B_q = \frac{\Gamma \left( {\frac{1}{1 - q}}
\right)}{\left( {\frac{1}{1 - q}} \right)^{1 \mathord{\left/
{\vphantom {1 2}} \right. \kern-\nulldelimiterspace} 2}\Gamma \left(
{\frac{1}{1 - q} - \frac{1}{2}} \right)}, \quad
 - 1 < q \le 1
\end{equation}

\noindent and

\begin{equation}
\label{eq27} B_q = \frac{1 + q}{2}\frac{\left( {\frac{1}{q - 1}}
\right)^{{ - 1} \mathord{\left/ {\vphantom {{ - 1} 2}} \right.
\kern-\nulldelimiterspace} 2}\Gamma \left( {\frac{1}{q - 1} +
\frac{1}{2}} \right)}{\Gamma \left( {\frac{1}{q - 1}} \right)},
\quad q \ge 1
\end{equation}

\noindent unlike the classical Boltzmann-Gibbs statistics. It is the
reason why the results obtained by J. A. S. Lima et al, are
problematic.

Substituting the marginal distribution Eqs.(\ref{eq21}) and
(\ref{eq24}) into the dielectric function Eq.(\ref{eq17}), we obtain

\begin{equation}
\label{eq28} \varepsilon _k^l = 1 + \frac{\omega _{pe}^2 }{k^2v_T^2
}\left[ {\frac{3q - 1}{2} - Z_q \left( x \right)} \right],
\end{equation}

\noindent where $\omega _{pe} = \sqrt {{4\pi n_0 e^2}
\mathord{\left/ {\vphantom {{4\pi n_0 e^2} m}} \right.
\kern-\nulldelimiterspace} m} $ is the plasma frequency, $x$ is the
dimensionless parameter, namely, $x = \omega \mathord{\left/
{\vphantom {\omega {\sqrt 2 }}} \right. \kern-\nulldelimiterspace}
{\sqrt 2 }kv_T $. $Z_q \left( x \right)$ is the generalized plasma
dispersion function in the context of Tsallis statistics,

\begin{equation}
\label{eq29} Z_q \left( x \right) = L_q \frac{x}{\sqrt \pi }\int
{\frac{1}{x - \xi + i\delta }\left[ {1 - \left( {q - 1} \right)\xi
^2} \right]^{\frac{1}{q - 1}}d\xi }
\end{equation}

\noindent where $\xi = {v_x } \mathord{\left/ {\vphantom {{v_x }
{\sqrt 2 }}} \right. \kern-\nulldelimiterspace} {\sqrt 2 }v_T $, in
the limit $q \to 1$, it is reduced to the standard form in B-G
statistics [11]

\begin{equation}
\label{eq30} Z_{q = 1} \left( x \right) = \frac{x}{\sqrt \pi }\int
{\frac{1}{x - \xi + i\delta }\exp \left( { - \xi ^2} \right)d\xi } .
\end{equation}

Using the Plemelj formula [11]

\begin{equation}
\label{eq31} \frac{1}{z\pm i0} = \wp \frac{1}{z} \mp i\pi \delta
(z),
\end{equation}

\noindent where $\wp $ denotes the principal value, then the
generalized plasma dispersion function Eq.(\ref{eq29}) can be
written as

\[
Z_q \left( x \right) = L_q \frac{x}{\sqrt \pi }\wp \int {\frac{1}{x
- \xi }\left[ {1 - \left( {q - 1} \right)\xi ^2} \right]^{\frac{1}{q
- 1}}d\xi - }
\]

\begin{equation}
\label{eq32} iL_q \sqrt \pi x\left[ {1 - \left( {q - 1} \right)x^2}
\right]^{\frac{1}{q - 1}},
\end{equation}

\noindent when $\omega \gg kv_T $, namely $x \gg 1$, the real part
of Eq.(\ref{eq32}) becomes

\[
L_q \frac{x}{\sqrt \pi }\wp \int {\frac{1}{x - \xi }\left[ {1 -
\left( {q - 1} \right)\xi ^2} \right]^{\frac{1}{q - 1}}d\xi }
\]

\begin{equation}
\label{eq33}
 = \frac{L_q }{\sqrt \pi }\wp \int {\left[ {1 - \left( {q - 1} \right)\xi
^2} \right]^{\frac{1}{q - 1}}\left( {1 + \frac{\xi }{x} + \frac{\xi
^2}{x^2} + \cdots } \right)d\xi } ,
\end{equation}

\noindent thus Eq.(\ref{eq32}) can be expressed as

\[
Z_q \left( x \right) \approx \frac{3q - 1}{2} + \frac{1}{2x^2} +
\frac{2}{5q - 3}\frac{3}{4x^4} -
\]

\begin{equation}
\label{eq34} iL_q \sqrt \pi x\left[ {1 - \left( {q - 1} \right)x^2}
\right]^{\frac{1}{q - 1}}.
\end{equation}

When $\omega \ll kv_T $, namely $x \ll 1$, introducing the
transformation $\xi = \eta + x$, then the real part of
Eq.(\ref{eq32}) can be written as

\[
L_q \frac{x}{\sqrt \pi }\wp \int {\frac{1}{x - \xi }\left[ {1 -
\left( {q - 1} \right)\xi ^2} \right]^{\frac{1}{q - 1}}d\xi } ,
\]

\[
 = L_q \frac{x}{\sqrt \pi }\wp \int {\left[ {1 - \left( {q - 1}
\right)\left( {\eta ^2 + 2\eta x + x^2} \right)} \right]^{\frac{1}{q
- 1}}\frac{d\eta }{ - \eta }}
\]

\begin{equation}
\label{eq35}
 \approx - L_q \frac{x}{\sqrt \pi }\wp \int {\left[ {1 - \left( {q - 1}
\right)\eta ^2} \right]^{\frac{1}{q - 1}}\frac{d\eta }{\eta }} = 0,
\end{equation}

\noindent then Eq.(\ref{eq32}) can be expressed as

\begin{equation}
\label{eq36} Z_q \left( x \right) \approx - iL_q \sqrt \pi x\left[
{1 - \left( {q - 1} \right)x^2} \right]^{\frac{1}{q - 1}},
\end{equation}

\noindent which can be used in investigating the low-frequency wave,
such as the ion acoustic waves. It should be noted that the process
is not pinpoint in Eq.(\ref{eq35}), the real part should be a very
small quantity, which may be obtained by numerical method. However,
the small quantity can be neglect when Eq.(\ref{eq32}) Substituted
into the dielectric function Eq.(\ref{eq28}).

Substituting Eq.(\ref{eq34}) into the dielectric function
Eq.(\ref{eq28}), according to the longitudinal dispersion relation
$Re\varepsilon _k^l = 0$, thus the generalized dispersion relation
of Langmuir wave is obtained,

\[
\omega ^2 = \omega _{pe}^2 + \frac{2}{5q - 3}3k^2v_T^2
\]

\begin{equation}
\label{eq37}
 = \omega _{pe}^2 + 3k^2v_{Tq}^2 ,
\end{equation}

\noindent where $v_{Tq} = \sqrt {{k_B T_q } \mathord{\left/
{\vphantom {{k_B T_q } m}} \right. \kern-\nulldelimiterspace} m} $
is the physical thermal speed, $T_q $ is the physical temperature
defined in Section 2. As expected, in the limit $q \to 1$,
Eq.(\ref{eq37}) reduces to

\begin{equation}
\label{eq38} \omega ^2 = \omega _{pe}^2 + 3k^2v_T^2 ,
\end{equation}

\noindent being the standard result in B-G statistics [11]. Thus the
dispersion relation of Langmuir wave obtained by J. A. S. Lima et al
[8]

\[
\omega ^2 = \omega _{pe}^2 + \frac{2}{3q - 1}3k^2v_T^2 ,
\]

\noindent is problematic. According to Fig.1 in their paper, we can
see that the dispersion relation for Tsallis formalism presents a
good fit to the experimental data when $0.7 < q < 0.85$, obviously,
it should be $0.82 < q < 0.91$ based on the correct result.

Next we will derive the expression of Landau damping. The Landau
damping rate can be written as [11]

\begin{equation}
\label{eq39} \gamma _k^l = \left. { - \frac{Im\varepsilon _k^l
}{\frac{\partial }{\partial \omega }Re\varepsilon _k^l }}
\right|_{\omega = \omega ^l} ,
\end{equation}

\noindent according to Eqs.(\ref{eq37}), (\ref{eq28}) and
(\ref{eq34}), we have that $\omega \approx \omega _{pe} $,
$Re\varepsilon _k^l \approx 1 - {\omega _{pe}^2 } \mathord{\left/
{\vphantom {{\omega _{pe}^2 } {\omega ^2}}} \right.
\kern-\nulldelimiterspace} {\omega ^2}$, $\left( {\partial
\mathord{\left/ {\vphantom {\partial {\partial \omega }}} \right.
\kern-\nulldelimiterspace} {\partial \omega }} \right)Re\varepsilon
_k^l = 2 \mathord{\left/ {\vphantom {2 {\omega _{pe} }}} \right.
\kern-\nulldelimiterspace} {\omega _{pe} }$, $Im\varepsilon _k^l =
L_q \sqrt {\pi \mathord{\left/ {\vphantom {\pi 2}} \right.
\kern-\nulldelimiterspace} 2} {\left( {\omega \omega _{pe}^2 }
\right)} \mathord{\left/ {\vphantom {{\left( {\omega \omega _{pe}^2
} \right)} {\left( {kv_T } \right)^3}}} \right.
\kern-\nulldelimiterspace} {\left( {kv_T } \right)^3} \cdot \left[
{1 - \left( {q - 1} \right){\omega ^2} \mathord{\left/ {\vphantom
{{\omega ^2} {\left( {2k^2v_T^2 } \right)}}} \right.
\kern-\nulldelimiterspace} {\left( {2k^2v_T^2 } \right)}} \right]^{1
\mathord{\left/ {\vphantom {1 {\left( {q - 1} \right)}}} \right.
\kern-\nulldelimiterspace} {\left( {q - 1} \right)}}$, combined with
Eq.(\ref{eq39}), we obtain the generalized Landau damping as

\[
\gamma _k^l = - L_q \sqrt {\frac{\pi }{8}} \omega _{pe} \left(
{\frac{k_d }{k}} \right)^3\cdot \
\]

\begin{equation}
\label{eq40} \left[ {1 - \left( {q - 1} \right)\left( {\frac{k_d^2
}{2k^2} + \frac{3}{5q - 3}} \right)} \right]^{\frac{1}{q - 1}},
\end{equation}

\noindent where $k_d = {\omega _{pe} } \mathord{\left/ {\vphantom
{{\omega _{pe} } {v_T }}} \right. \kern-\nulldelimiterspace} {v_T }$
is the electronic Debye wave number. In the limit $q \to 1$,
Eq.(\ref{eq40}) reduces to

\begin{equation}
\label{eq41} \gamma _k^l = - \sqrt {\frac{\pi }{8}} \omega _{pe}
\left( {\frac{k_d }{k}} \right)^3\exp \left( { - \frac{k_d^2 }{2k^2}
- \frac{3}{2}} \right),
\end{equation}

\noindent which is the classical Landau expression for the damping
decrement in the framework of B-G statistics [11].

\section{Summary}
\label{sect-res} In this comments, we have discussed the dispersion
property and Landau damping of Langmuir wave in an unmagnetized,
collisionless, isotropic plasma with the nonextensive distribution
in Tsallis statistics. The correct generalized dispersion relation
and Landau damping are obtained. In the limiting case ($q \to 1)$
the classical results based on the B-G statistics are recovered. It
is our hope that the discussion here will serve as a useful
introduction to the field of plasma physics.

\vspace*{-2ex}
\begin{acknowledgments}
\vspace*{-2ex} The work was supported by the National Natural
Science Foundation of China under the grant Nos. 10963002, the
International S{\&}T Cooperation Program of China (2009DFA02320) and
Jiangxi Province, Program for Innovative Research Team in Nanchang
University, and the National Basic Research Program of China (973
Program) (No. 2010CB635112).
\end{acknowledgments}


\end{document}